\documentclass[prd,aps,amsmath,amssymb,nofootinbib,preprintnumbers]
{revtex4}

\usepackage{graphicx}
\usepackage{dcolumn}
\usepackage{bm}
\usepackage{amsmath}
\usepackage{amsfonts}
\usepackage{ifthen}
\usepackage{psfrag}

\def\ls{\mathrel{\lower4pt\vbox{\lineskip=0pt\baselineskip=0pt
           \hbox{$<$}\hbox{$\sim$}}}}
\def\gs{\mathrel{\lower4pt\vbox{\lineskip=0pt\baselineskip=0pt
           \hbox{$>$}\hbox{$\sim$}}}}
\def\drawbox#1#2{\hrule height#2pt

\hbox{\vrule width#2pt height#1pt \kern#1pt
              \vrule width#2pt}
              \hrule height#2pt}

\def\Asym#1#2{\vcenter{\vbox{\drawbox{#1}{#2}
              \kern-#2pt       
              \drawbox{#1}{#2}}}}

\newcommand{\be}{\begin{equation}}
\newcommand{\ee}{\end{equation}}
\newcommand{\bea}{\begin{eqnarray}}
\newcommand{\eea}{\end{eqnarray}}

\begin{document}

\title {\Large Schizophrenic Neutrinos and  $\nu$-less Double Beta Decay}
\author{\bf Rouzbeh Allahverdi$^{1}$, Bhaskar Dutta$^{2}$ and Rabindra N. Mohapatra$^{3}$}
\affiliation{$^{1}$~Department of Physics and Astronomy, University of New Mexico, Albuquerque, NM 87131, USA \\
$^{2}$~Department of Physics and Astronomy, Texas A\&M University, College Station, TX 77843-4242, USA \\
$^{3}$~Maryland Center for Fundamental Physics, Department of Physics, University of Maryland, College Park, MD 20742, USA}

\begin{abstract}
We point out a new possibility for neutrinos where all neutrino
flavors can be part Dirac and part Majorana. Our primary
motivation comes from an attempt to use supersymmetric see-saw
models to tie inflation, baryon asymmetry of the Universe and dark
matter to the neutrino sector. The idea however could stand on its
own, with or without supersymmetry. We present a realization of
this possibility within an $S_3$ family symmetry for neutrino
masses, where we obtain tri-bi-maximal mixing for neutrinos to the
leading order. The model predicts that for the case of inverted
hierarchy, the lower limit on the neutrino mass measured in
neutrinoless double beta decay experiments is about a factor of
two larger than the usual Majorana case.
\end{abstract}

\begin{flushright}
MIFPA-10-36,  UMD-PP-10-011
\end{flushright}
August 2010

\maketitle

\section{Introduction}

Experiments over the past two decades have conclusively
established that neutrinos have mass. The true nature of the
neutrino mass however is unknown since available observations
based on flavor oscillations do not tell us whether it is its own
anti-particle (Majorana type) or not (Dirac type). Unlike the
quarks and charged leptons, both these possibilities are allowed
for the neutrinos since they are electrically neutral. Numerous
neutrinoless double beta decay experiments are under way or in
preparation to settle this question.

An intermediate possibility that has been discussed in literature
is known as the pseudo-Dirac case\cite{pseudo} where one includes
a very tiny amount of the Majorana mass for each neutrino flavor
which has Dirac type mass. The Majorana entry in this
case must be very tiny ($\leq 10^{-10}$ eV) in order to be
consistent with current solar neutrino observations\cite{andre}.
In this note we point out a new class of  possibilities where each
neutrino flavor can have a large admixture of both Dirac and Majorana
masses under certain circumstances. We point out the experimental
implications of this possibility as well as its possible
theoretical origin.

While discussing the Dirac versus Majorana nature of neutrinos, it
is usual to frame the discussion in terms of the neutrino flavor
eigenstates that are emitted in beta decay or other weak interaction processes.
 When the neutrinos
travel in free space, however, they do so as mass eigenstates, which are
linear superpositions of the flavor eigenstates. This fact
is responsible for neutrino oscillation phenomena.  In this note
we point out that the possibility of one of the neutrino mass
eigenstates having a Dirac mass at the tree level with the others
having Majorana type mass, appears consistent with all current
observations. Since in this case, each flavor eigenstate is a
large admixture of both Dirac and Majorana masses, we call this
``schizophrenic neutrino" alternative. This is different from the usual pseudo-Dirac
cases discussed in literature where the Majorana admixture
 is tiny compared to the Dirac mass whereas in our case Majorana admixture
  is as large or larger than the Dirac mass
for each flavor. This possibility implies
 distinct predictions for neutrinoless double beta decay
searches compared to the case where the neutrinos are pure
Majorana type and could be used to test the schizophrenic hypothesis.

On the theoretical side, the mass eigenstate having the Dirac mass
must have a Dirac Yukawa coupling which is extremely tiny ($\sim
10^{-12}$) whereas the other masses could arise from high mass
scale physics as in seesaw models~\cite{seesaw} with much larger Yukawa couplings. The Dirac type
mass eigenstate would pair up with a right-handed (RH or sterile) neutrino
$(\nu_s$) to form the Dirac mass.  A priori, we do not know which
of the three mass eigenstates has the Dirac nature. In this paper,
we consider a specific model where we want to get tri-bi-maximal
pattern~\cite{tbm} for the PMNS matrix. This suggests that the
eigenstates  be representations of an $S_3$ symmetry group. The
model then picks the ``middle'' eigenstate $\nu_2$ (the one that
determines solar neutrino oscillations) as Dirac type since this is an $S_3$ singlet
whereas the other two are Majorana. In a generic model, any or in fact any two of the
mass eigenstates could have Dirac type mass.

While one would like to understand the small Dirac Yukawa coupling
as a consequence of some high scale theory, it is comforting to
know that it stable under radiative corrections due to chiral
symmetry (or in this case under the symmetry $\nu_s\to -\nu_s$).
There may be other motivations for the existence of such tiny
Yukawa couplings. One such motivation in supersymmetric versions
of such models comes from attempts to use the RH sneutrino to
drive inflation~\cite{akm}. In such a scenario, small Dirac
coupling is essential to make the inflation predictions for
density fluctuations consistent with observations.
 However the hypothesis of schizophrenic neutrinos could be considered independently of
this. As indicated earlier, a testable prediction of this model is
that in the case of inverted hierarchy, the lower bound on
$m_{\beta \beta}$ measured in neutrinoless double beta decay
searches is roughly twice that of usual inverted hierarchy models
in literature. This model will therefore be easier to rule out by
the current generation of $\beta \beta_{0 \nu}$ experiments if
long base line oscillation searches indicate inverted neutrino
mass ordering.


We hasten to point out that this kind of pattern for neutrino
masses is not protected by a symmetry. As a result, when loop
corrections are taken into account, tiny corrections of order
$g^2m^2_\tau/(32\sqrt{6} \pi^2 M^2_W) \sim 4\times 10^{-7}$ appear
giving the Dirac eigenstate a pseudo-Dirac mass splitting of order
$10^{-14}$ eV. These corrections have no impact on our prediction
for $\beta \beta_{0 \nu}$ decay.

\section{Motivation from cosmology}
In this section, we review the cosmological motivation for small
neutrino Dirac coupling in a supersymmetric seesaw model. We
consider a  supersymmetric extension of MSSM based on the gauge
group $SU(2)_L \times U(1)_{I_{3R}} \times U(1)_{B-L}$ which
requires that there be three RH neutrinos $N$ ($\equiv \nu^c$ in
SUSY language) to cancel anomalies (with the eventual possibility
of SO(10) grand unification).
A combination of superpartners $\phi \equiv (\tilde N + H_u +
\tilde L/\sqrt{3})$ in this theory is a $D$-flat directions under
the whole gauge symmetry
, and is also $F$-flat when
neutrino Yukawa couplings $h$ are turned off.
As shown in~\cite{akm}, this flat direction
can act as the inflaton.
The neutrino Yukawa couplings in
combination with the soft mass and $A$-term for $\phi$
leads to a potential for the radial component of
$\phi$ (denoted as $\varphi$) of the form~\cite{akm}
\begin{eqnarray} \label{potential}
V(\varphi )~=~ {m^2_\phi \over 2} \varphi^2 + {h^2 \over 12} \varphi^4 - {A h \over 6 \sqrt{3}} \varphi^3
\end{eqnarray}
where $m^2 = (m^2_{{\tilde N}_0} + m^2_{H_u} + m^2_{\tilde L})/3$
 can lead to inflection point inflation~\cite{akm}
and the amplitude
of observationally relevant density perturbations (as measured by
COBE and WMAP)
%
%
matches the observed value $\delta_H \sim 1.9 \times 10^{-5}$ for
weak scale masses $m_\phi \sim {\cal O}(100)$ GeV, provided that
$h \sim 10^{-12}$ (for details, see~\cite{akm}). The
neutrino mass is intimately connected to $h$. For instance, if
neutrinos are Dirac fermions, we have $m_\nu = h \langle H_u
\rangle$, where $\langle H_u \rangle = (174 ~ {\rm GeV}) ~ {\rm
sin} \beta$ and ${\rm tan} \beta$ is the ratio of vacuum
expectation value (VEV) of Higgs fields of the minimal
supersymmetric standard model (MSSM). Then $h \sim 10^{-12}$ would
give rise to $m_\nu \sim {\cal O}(0.1)$ eV, which is precisely in
the range of interest for neutrino oscillations. We could take
this as a hidden message from cosmology that at least one of the
neutrinos can be dominantly of Dirac nature, and study its
implications for neutrino masses and mixings.

It is important to emphasize that the inflation model constrains
the coupling of only one of the RH neutrinos whose superpartner is
responsible for inflation. That RH neutrino could be the Dirac
partner of one of the light neutrino combinations making it a
Dirac neutrino. The other two RH neutrinos have unconstrained
Yukawa couplings that take natural values $(\sim 10^{-5}-0.1)$,
and hence their mass must be heavy.
Note however that the heavy RH neutrinos must not mix with the RH
neutrino whose superpratner is part of the inflaton so as not to
spoil the picture of inflation mentioned above. The simplest
possibility for neutrino masses in this case would therefore
appear to be that one linear combination of the flavor eigenstates
is a Dirac fermion whereas the other two will be Majorana and get
their mass via the see-saw mechanism. Below we suggest this as new
picture for neutrino masses.


\section{An $S_3$ model for schizophrenic neutrinos}

One of the challenges in neutrino mass physics is to understand
the observed near tri-bi-maximal mixing pattern among different
flavors. Discrete symmetries have been discussed extensively as a
way to address this issue~\cite{discrete} and the group $S_3$ is
one of the symmetries that appears promising in this context and
we use it in our discussion in this paper. The basic assumptions of our
neutrino model can therefore be summarized as follows:

\begin{itemize}

\item The extended gauge group responsible for neutrino masses
consists of a local $B-L$ symmetry~\cite{marshak,BL}, which
requires the existence of three RH neutrinos for anomaly
cancellation.

\item  One of the RH neutrinos (whose superpartner is the inflaton
field) couples to a linear combination of all neutrino flavors
with a Yukawa coupling of order $10^{-12}$ so that it gets a Dirac
mass without any need for see-saw, whereas the remaining
orthogonal combinations get their masses from the conventional
see-saw mechanism.

\item The three standard model lepton doublets transform into one another
under a flavor $S_3$ discrete symmetry.

\end{itemize}

The first assumption is quite well motivated and has been widely
discussed in literature. It also naturally incorporates ${\tilde
N}$ along with $H_u$ and ${\tilde L}$ into a single $D$-flat
direction that can drive inflection point inflation. The second
assumptions is motivated by the cosmological scenario discussed
above.

As already mentioned, our neutrino model is based on the idea that
only one of the neutrino flavor combinations corresponding to a
mass eigenstate has a small Yukawa coupling to one RH neutrino
whereas the other two combinations get their masses from the
seesaw mechanism. If we take the tri-bi-maximal matrix as the
leading order PMNS matrix, then one might start thinking of a
discrete symmetry group which has one singlet and one doublet as
part of its irreducible representations and the singlet one being
the Dirac neutrino whereas the doublet combinations becoming
Majorana. One such example used in literature is the $S_3$
group~\cite{S3} which proves convenient for our discussion.

We assume the $S_3$ to permute the three families of leptons
$(L_e, L_\mu, L_\tau)$ among themselves. Of course, it is well
known that this is a reducible representation of $S_3$ group but
the following linear combinations of these fields transform as
singlet and two dimensional representations of $S_3$:
\begin{eqnarray}
&& L_2 ~=~ \frac{1}{\sqrt{3}}(L_e + L_\mu + L_\tau) ~ ~ ~ ~ ~ ({\rm Singlet}) \, \nonumber \\
&& L_1 ~=~ \frac{1}{\sqrt{6}}(2L_e - L_\mu - L_\tau)  ~ ~ ~ ({\rm Doublet}) \, \nonumber \\
&& L_3 ~=~ \frac{1}{\sqrt{2}}(L_\mu - L_\tau)  ~ ~ ~ ~ ~ ~ ~ ~ ~ ~ ~ ({\rm Doublet}) \, .
\end{eqnarray}
We assume that muon type RH neutrino is the $S_3$ singlet whereas
$(N_e, N_\tau)$ form a doublet. The masses of these doublet RH
neutrinos can be chosen different by appropriate choice of
symmetry breaking (see below). The effective lepton Yukawa
coupling after integrating out $N_e$ and $N_\tau$ can then be
written as
\begin{eqnarray}\label{lagrangian}
{\cal L}_\nu &~=~& h L_2 H_u N_\mu~+~\frac{h^{2}_1} {M_{N_e}}(L_1 H_u)^2+
\frac{h^{2}_3}{M_{N_\tau}}(L_3 H_u)^2 \, \nonumber \\
&~+~& {\rm h.c.} \,
\end{eqnarray}
After the electroweak symmetry breaking, this gives rise to one Dirac
neutrino corresponding to the mass eigenstate $\nu_2$ and two
Majorana eigenstates $\nu_1, \nu_3$ and clearly leads to
tri-bi-maximal form for the PMNS matrix provided the charged
lepton mass matrix is diagonal.

The effective Lagrangian in \ref{lagrangian} could for instance
arise in an $SU(2)_L\times U(1)_{I_{3R}}\times U(1)_{B-L}$ theory
supplemented by a global discrete symmetry $S_3\times D$, (where
$D$ is a product of extra $Z_n$ symmetries) if we take an $S_3$
doublet Higgs fields $(\Delta^c_1, \Delta^c_2)$ and a singlet
field $\Delta^c_0$ (all with $B-L$ charge $-2$ and $I_{3R}$ charge
+1) with VEVs $\langle \Delta_1 \rangle =0$ and others with
non-zero VEV. This will generate different Majorana masses
$M_{N_e}$ and $M_{N_\tau}$ for the $S_3$ doublet RH neutrinos.

In this model, inflation occurs along the flat direction
corresponding to the first superpotential term in
Eq.~\ref{lagrangian}. The coupling between $N_\mu$ and $(N_e,
N_\tau)$ can be forbidden by e.g., a $Z_8$ symmetry contained in
$D$ under which $N_\mu\to -i N_\mu$ and a gauge singlet field $X$
with $X \to e^{i \pi/4}X$ with all other fields invariant. The
Dirac coupling of $N_\mu$ can be obtained from a higher
dimensional coupling $(\lambda L_2 H_u N_\mu X^2/M_{Pl}^2)$, where
$\langle X \rangle \sim 10^{12}$ GeV or so. At the inflection
point VEV ($\sim 10^{12}$ GeV~\cite{akm}), this interaction
then yields the effective Dirac coupling of $N_\mu$ in
Eq.~\ref{lagrangian}. An additional RH neutrino mixing term
$(N_\mu N_{e,\tau}\Delta^c_{1,2}X^2/M_{Pl}^2)$ is allowed by the
$Z_8$ symmetry, but has negligible contribution to masses and
mixings and can be ignored. We need to add the fields
$(\overline\Delta^c_1, \overline\Delta^c_2)$ and a singlet field
$\overline\Delta^c_0$ to preserve supersymmetry below the $B-L$
scale as well as to cancel anomalies.

Turning to the charged lepton mass matrix, neutrino oscillation
data require that it be nearly diagonal. We employ the technique
used in the second reference in~\cite{S3}. We add three gauge
singlet superfields $(\sigma_e, \sigma_\mu, \sigma_\tau)$ and
three extra $Z_n$ symmetries, i.e. $Z_{n,e}\times Z_{n,\mu}\times
Z_{n,\tau}$, with RH lepton fields $e^c, \mu^c, \tau^c$
transforming as $\omega^p_{e,\mu,\tau}$ and singlet fields
transforming as $\omega^{-p}_{e,\mu, \tau}$. Both sets of three
fields also transform under $S_3$ exactly like the lepton doublet
fields. We can write down the corresponding Yukawa superpotential
as
%
\begin{eqnarray}
{\cal W}_{l, Y}~=~\frac{1}{M} h_e H_d~(L_e \sigma_e e^c~+~L_\mu \sigma_\mu
\mu^c~+~L_\tau \sigma_\tau \tau^c) .
\end{eqnarray}
There can be another term where the $L_e, L_\mu, L_\tau$ are
permuted among themselves. This will contribute to the
off-diagonal elements of the charged lepton mass matrix after
symmetry breaking. We set this coupling to zero. Now by adjusting
the VEVs of the singlet fields, we can get diagonal mass matrix
for the charged leptons. On the other hand, if the small
contributions to the mass matrix coming from the permuted terms
are kept, there will corrections to the tri-bi-maximal form e.g.
it will lead to non-zero $\theta_{13}$.

\section{Implications for neutrinoless double beta decay}

This neutrino mass model has an interesting implication for
neutrinoless double beta decay. Recall that in the conventional
all Majorana neutrino case, the light neutrino contribution to
$\beta \beta_{0\nu}$ decay takes the form
%
$m_{\beta \beta}~=~\sum_{i} U^2_{e i} m_i$, $(i=1,2,3)$, where $U_{e i}$ are entries of the PMNS matrix.
%
In the inverted hierarchy scenario, this leads to the following
lower bound  for the conventional three Majorana neutrino
case~\cite{vissani}
\begin{eqnarray} \label{inverted1}
\vert m_{\beta \beta} \vert \simeq |({\rm cos}^2 \theta_{\odot} +
e^{i \alpha} {\rm sin}^2\theta_{\odot}) m_{\rm atm}|
 \geq \frac{m_{\rm atm}}{3} \approx 17~{\rm meV} && \, \nonumber \\
({\rm Conventional}). && \,
\end{eqnarray}
In our model, however, the second neutrino mass eigenstate is a
Dirac type state and therefore has no contribution to $\beta
\beta_{0 \nu}$ decay.
This leads to the following lower bound for inverted case:
\begin{eqnarray} \label{inverted2}\vspace{-1 cm}
\vert m_{\beta \beta} \vert \simeq {\rm cos}^2\theta_{\odot}
m_{\rm atm} \geq \frac{2m_{\rm atm}}{3} \approx 34 ~
{\rm meV} && \, \nonumber \\
({\rm Dual}), && \,
\end{eqnarray}
%
which is roughly twice the value of the conventional
case~(\ref{inverted1}). This makes it easier to rule out our model
in the current generation of neutrinoless double beta decay
searches, provided we have independent evidence, e.g. from long
base line neutrino experiments for inverted mass hierarchy.

In the normal hierarchy scenario, the corresponding formula becomes
$m_{\beta \beta} \simeq (m_1 {\rm cos}^2 \theta_{\odot} + e^{i \alpha^{\prime}}
{\rm sin}^2 \theta_{13} m_{\rm atm})$, which is different from the
conventional three Majorana case. The precise value in this case
depends on the unknown Majorana mass of $\nu_1$ as well as the
value of $\theta_{13}$.

\begin{figure}
\begin{center}
\includegraphics[width=.48\textwidth]{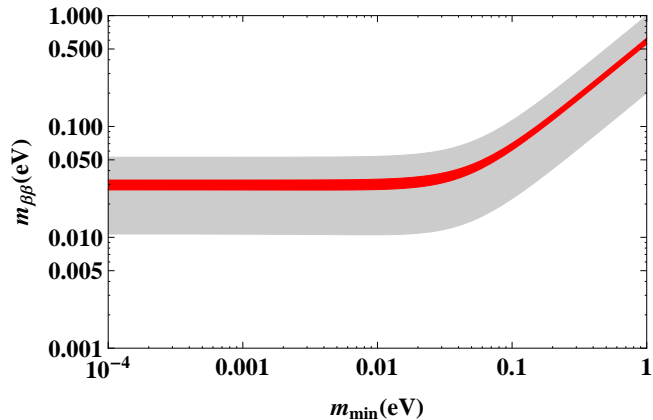}
\end{center}
\vskip -0.1in \caption{We plot $|m_{\beta \beta}|$ as a function
of the lightest neutrino mass for the case of inverted hierarchy.
The red (dark) band shows the prediction from the two Majorana and
one Dirac neutrino scenario and the gray shaded region shows the
conventional three Majorana neutrino case. } \label{doublebeta}
\end{figure}

In Fig.~\ref{doublebeta}, we plot $|m_{\beta \beta}|$ as a
function of the lightest neutrino mass min($m_j) = m_{min}$ (which sets the absolute neutrino mass scale in the case of
inverted hierarchy). The red (dark) band shows the prediction of
our scenario and the gray shaded region shows the usual three
Majorana neutrino scenario in the case of inverted hierarchy. The
masses and mixing angles used for the figure are as
follows~\cite{GonzalezGarcia:2010er}: $\Delta
m^2_{\odot}$=$(7.59\pm0.20^{+0.61}_{-0.69}) \times 10^{-5}$eV$^2$,
$\Delta m^2_{\rm atm}$ = $(2.46 \pm 0.12 {\pm 0.37}) \times
10^{-3}$eV$^2$, $\theta_{13}<12.5^o$, $\theta_{\odot}={34.4^{o}
\pm 1^{+3.2}_{-2.9}}$ and $\theta_{\rm atm}=42.8^{o ~ +4.7
+10.7}_{~ ~ -2.9 -7.3}$. We can see that in the case of inverted hierarchy (corresponding to $m_{min} < 0.05$ eV) the lower limit on $\vert m_{\beta \beta} \vert$ measured in neutrinoless double beta decay
experiments is about a factor of two larger than the conventional
case.

\section{Comments}

We now make some comments on the model
described here.

(i) Since the Dirac nature of the second neutrino mass eigenstate
is not protected by any symmetry, radiative corrections will
induce Majorana component to its mass. The self-energy corrections
to $\nu_i$ masses due to $W^+ \ell^-$ intermediate states will
lead to kinetic mixings between the different mass eigenstates
that depends on the charged lepton masses: $\epsilon_{ij} \sim
(\sum_{k} {U_{ik} U^*_{kj}g^2 m^2_{\ell_k}}{32\sqrt{6} \pi^2
M^2_W})$. This mixing is of order $4\times 10^{-7}$. When the
kinetic energy term in the Lagrangian is diagonalized, this leads
to mixing terms (for the normal hierarchy case), e.g. $m_\delta
\nu_3 \nu_2 + ...$, where $m_\delta \sim m_{\nu_3} \epsilon_{23} +
...$. This introduces a Majorana mass term $\delta m_2 \nu_2
\nu_2$ with $\delta m_2 \sim 10^{-14}$ eV. It, being very small,
keeps the Dirac nature of $\nu_2$ to very high precision. This is
also consistent with a bound $\sim 10^{-9}$ eV on this parameter
from solar neutrino observations~\cite{andre}. The same result
holds for the inverted hiercrhy case with $\nu_1$ and $\nu_3$
interchanged. In the SUSY version, quantum corrections that mix
the slepton states introduce a Majorana component for the Dirac
neutrino.
At one loop this effect results in $\delta m^2_{ij} \sim [(Y^{\dagger}_\ell Y_\ell)_{ij}/16\pi^2]~m^2_0~{\rm ln} ({M^2}/{M^2_Z})$,
which is of the same order as that mentioned before.

(ii) We wish to emphasize that our scenario is different from the
usual pseudo-Dirac scenario~\cite{pseudo} discussed in the
literature. Our light neutrino mass matrix is a $4 \times 4$
matrix such that one of its eigenstates forms a Dirac pair with
the sterile neutrino and other two eigenstates are Majorana. The
Dirac eigenstate gives rise to a pseudo-Dirac pair only at the
one-loop level.

(iii) The masses of the two heavy RH neutrinos depend on the scale
at which $B-L$ is broken, and can be as low as ${\cal O}(1)~{\rm
TeV}$. The decay of heavy Majorana neutrinos, and their SUSY
partners, can generate baryon asymmetry of the Universe. If
$M_{N_e}, M_{N_\tau}$ are of order TeV, resonant leptogenesis will
be a relevant solution. However, in the $S_3$ symmetric model,
this does not work since it will require the first and the third
neutrino masses be almost equal. The oscillation data will be hard
to fit with this pattern. However, soft leptogenesis~\cite{soft}
can work well in the model for a wide range of Majorana masses.

(iv) In this model, either the MSSM neutralino or the superpartner
of the RH component of the Dirac neutrino can play the role of
dark matter. The latter is naturally the lightest of the RH
sneutrinos since its mass receives contribution from SUSY breaking
alone. If the $B-L$ is broken around TeV, it can obtain the
correct relic density via thermal freeze out~\cite{akm}.
This also makes the corresponding $Z^{\prime}$ accessible at the LHC. On the other hand,
for a high scale $B-L$ the usual MSSM neutralino is a good dark matter candidate. The role of B-L in this
case is to provide the R-parity symmetry naturally.

(v) The impact of the RH neutrino, which is responsible for the
Dirac mass, on Big Bang Nucleosynthesis also depends on the scale
at which $B-L$ is broken. For example, for $M_{Z'} \sim 10$ TeV,
the RH neutrinos decouple at $T_D \sim 1$ GeV, while for $M_{Z'}
\sim 1 $ TeV we have $T_D \sim 100$ MeV. In the latter case this
amounts to $N^{\rm eff}_\nu \simeq 4$, whereas $N^{\rm eff}_\nu
\simeq 3.1$ in the former case. For a high scale $B-L$ the RH
neutrinos
decouple much earlier, and hence $N^{\rm eff}_\nu \approx 3$.\\
\\
In summary, motivated by cosmology, we have pointed out a new picture for neutrino masses
with the novel feature that one of the mass eigenstates is a Dirac fermion (at the tree-level) whereas the other two are
Majorana type. We presented an $S_3$ realization of this idea that
leads to tri-bi-maximal mixing for leptons in the leading order.
This model can be ruled out by the current generation of
neutrinoless double beta decay searches if inverted mass hierarchy
is indicated by long base line neutrino oscillation experiments
and neutrinoless double beta decay searches give
$|m_{\beta\beta}| \lesssim 34$ meV.

{\it Acknowledgement}: We wish to thank A. de Gouvea, S. T. Petcov
and L. Wolfenstein for useful discussions and helpful comments. BD
is supported  by the DOE grant DE-FG02-95ER40917. RNM is supported
by the NSF under grant PHY-PHY0968854.

\end{document}